\documentclass[prd,twocolumn,showpcs,amsmath,amssymb,nofootinbib,preprintnumbers,balancelastpage]{revtex4}
\pdfoutput=1
\usepackage{epsfig}
\usepackage{bbold}
\usepackage{amsmath}
\usepackage{bm}
\usepackage{times}
\usepackage{graphicx}
\usepackage{color}
\usepackage{slashed}
\usepackage{graphicx}
\usepackage{amsmath}
\usepackage{tikz}
\usetikzlibrary{positioning,shapes}
\usepackage{relsize,caption} 
\usepackage[latin1]{inputenc}
\usepackage{tikz}
\usetikzlibrary{trees}
\usetikzlibrary{decorations.pathmorphing}
\usetikzlibrary{decorations.markings}
\usetikzlibrary{positioning,arrows}
\usetikzlibrary{decorations.pathmorphing}
\usetikzlibrary{decorations.markings}
\usetikzlibrary{decorations.pathreplacing,calc}
\usetikzlibrary{decorations.pathmorphing,decorations.markings,trees,positioning,arrows}   
\newif\ifmirrorsemicircle

\def\bea{\begin{eqnarray}}
\def\eea{\end{eqnarray}}
\def\bean{\begin{equation*}}
\def\eean{\end{equation*}}

\begin{document}

\preprint{UCI-HEP-TR-2015-18}

\title{Dark Matter from Unification of Color and Baryon Number}

\author{Bartosz~Fornal}
\affiliation{\\ Department of Physics and Astronomy, University of California, Irvine, CA 92697, USA}
\author{Tim~M.~P.~Tait}
\affiliation{\\ Department of Physics and Astronomy, University of California, Irvine, CA 92697, USA}
\date{\today}

\begin{abstract}
We analyze a recently proposed extension of the Standard Model based on the 
$SU(4) \times SU(2)_L \times U(1)_X$
gauge group, in which baryon number is interpreted as the fourth color and dark matter emerges
as a neutral partner of the ordinary quarks under $SU(4)$.  
We show that under well-motivated minimal flavor-violating
assumptions the particle spectrum contains a heavy 
dark matter candidate which is dominantly the partner of the right-handed top quark.
Assuming a standard cosmology, the correct thermal relic density through freeze-out is obtained
for dark matter masses around 2 -- 3 TeV.
We examine the constraints and future prospects for direct and indirect searches for dark matter. We also
briefly discuss the LHC phenomenology, which is rich in top quark signatures, and investigate the prospects for discovery at a 100 
TeV hadron collider.
\vspace{11mm}
\end{abstract}

\maketitle

\section{Introduction}

The particle identity of the dark matter (DM) is among the most pressing questions confronting particle physics
today.  It is clear that DM requires an extension of the Standard Model (SM) and it is likely that an understanding of how DM fits into the context of the SM
will offer hints about the underlying structure which gave rise to it.

Among its most mysterious properties of DM is the fact that it is long-lived, either being exactly
stable or with a lifetime of order that of the Universe itself.  Its observational properties are
incompatible with it possessing either of the known exactly conserved gauge charges, and thus one
would naively expect it to decay quickly.  The fact that it is massive and yet (at least to very good
approximation) stable provides an important clue about its identity, suggesting that some kind of
symmetry is in operation.  It is tempting to postulate a connection with the accidental symmetries of the
SM, baryon and lepton number, as these are also thought to explain the surprisingly long
lifetime of the proton.
Indeed, the past years have seen an increase in interest of models gauging $U(1)$ baryon
motivated by the proton's lifetime, with early constructions \cite{Pais:1973mi, Rajpoot:1987yg, Foot:1989ts, Carone:1995pu, Georgi:1996ei} 
paving the way to more complete models \cite{FileviezPerez:2010gw}.  Already, this 
idea was found in some constructions to lead naturally
to theories including DM candidates \cite{Duerr:2013dza,Perez:2014qfa,Duerr:2014wra,Arnold:2013qja}.  

Extending this idea, in \cite{Fornal:2015boa} a model unifying gauged baryon number and color into a single non-Abelian gauge group was
constructed. The theory is based on the gauge symmetry 
$SU(4) \times SU(2)_L \times U(1)_X$, with the SM quarks forming quadruplets together
with new colorless ``quark partner" fields which obtain vector-like masses after $SU(4)$ breaking. 
The quark partners of the right-handed up-type quarks are SM gauge singlets and 
have suitable properties to play the role of DM.  If the mass of the lightest
of such states is chosen to be a few GeV and the gauge structure is
supplemented by additional UV interactions, a picture in which the DM number density
is determined by a primordial particle-anti-particle asymmetry connected to
the asymmetry in baryon number can be realized.

This asymmetric limit is interesting, but also has a few weak points.  The quark partners have their
own analogue of Yukawa interactions, and thus generically one would not expect the notion of flavor in both
the quark and quark partner sectors to be aligned, opening the door for large contributions to
flavor-changing neutral currents mediated by the GeV mass DM particle.  
The need for at least one of the quark partners to have a GeV scale mass precludes the invocation
of symmetry-based arguments such as minimal flavor violation (MFV) \cite{D'Ambrosio:2002ex}
as a remedy.
Perhaps even more unwieldy is the
need to introduce an additional sector of light states into which the DM can annihilate
(or live with extreme tuning of parameters)
to deplete its primordial
symmetric component, a generic issue for models of asymmetric DM \cite{Zurek:2013wia}.

These concerns are largely ameliorated if the DM is much heavier and its
density is symmetric, resulting from its interactions with the
SM quarks freezing out at much higher temperatures.  While no longer trying to motivate the observed
correspondence between the observed densities of DM and baryons in the Universe, such a limit
arises naturally when baryon number and color unify, without the need for ad hoc assumptions or additional
ingredients.  In the current work, we abandon the connection to the baryon asymmetry
and consider the $SU(4)$ model in the limit where all of the quark partners have masses on the order of the $SU(4)$ breaking
scale.  The $D$--$\overline{D}$ mixing constraints (derived in \cite{Golowich:2007ka} for structurally similar leptoquark models) suggest
that even in this limit flavor is generically a problem unless there is sufficient alignment between the quark and quark partner
Yukawa interactions.  As detailed below, we invoke MFV, which results in sufficient alignment for the the first two generations
such that $D$--$\overline{D}$ constraints allow for quark partners with TeV scale masses.
The result is a variant of models where the DM is ``top-flavored" 
 \cite{Kile,Batell,Kamenik,Agrawal,Kumar,Batell2,Agrawal2,Zhang:2012da}, leading to interesting and distinct phenomenology.

This paper is organized as follows: 
In Section~\ref{sec:model}, we review the unification of baryon number with color, focusing on the features most important for DM. 
In Section~\ref{sec:dm} we compute the rates of annihilation of the DM candidate, as well as its scattering with heavy nuclei, allowing us
to identify the regions in which we expect the correct relic density (assuming a standard cosmology) and constraints from the null searches
for direct detection of DM. 
Section~\ref{sec:lhc} is devoted to a brief review of the associated collider signals. We summarize our results in Section~\ref{sec:conclusions}.

\section{Unification of Color and Baryon Number}
\label{sec:model}

\noindent
In this section we provide a brief summary of the most important features of the model for DM and its interactions
(for more details see \cite{Fornal:2015boa}).
The underlying gauge structure is:
\begin{eqnarray}
SU(4) \times SU(2)_L \times U(1)_X \ ,
\label{group}
\end{eqnarray}
where $X$ is a linear combination of hypercharge and the diagonal $T^{15}$ generator of $SU(4)$.
The SM quarks are promoted to $SU(4)$ quadruplets: $\hat{Q}_{L}$, $\hat{u}_{R}$, and $\hat{d}_{R}$,
consisting of the ordinary quark triplets: $Q_L$, $u_R$, and $d_R$, and additional uncolored
$SU(4)$ partner fields: $\tilde{Q}_L$, $\tilde{u}_R$ and $\tilde{d}_R$.  A generational index should be
understood as implicit.

A phenomenologically viable, anomaly-free set of fields is given by: 
\bea
&&\hspace{-3mm}\hat{Q}_{L} = \ \left(4, 2, 0\right) , \  \ \, \, \ \ \ \hat{u}_{R} = \left(4, 1, \tfrac{1}{2}\right) , \ \  \ \hat{d}_{R} = \left(4, 1, -\tfrac{1}{2}\right)  ,  \nonumber\\
&&\hspace{-3mm}Q'_{R} = \left(1, 2, -\tfrac{1}{2}\right)  , \, \ \ \ u'_{L} = \left(1, 1, 0\right)  , \  \ \ \ d'_{L} = \left(1, 1, -1\right) ,\nonumber\\
&&\hspace{-3mm}\ l_{L} \ = \left(1, 2, -\tfrac{1}{2}\right)  , \, \ \ \ e_{R}  = \left(1, 1, -1\right) ,\nonumber\\
&&\hspace{-3mm} \ \hat{\Phi} \ \, = \ \left(4, 1, \tfrac{1}{2}\right)  ,  \ \ \ \ \,  \  H= \, \left(1, 2, \tfrac{1}{2}\right) ,
\eea
where the numbers in parenthesis indicate the representations of $SU(4)$, $SU(2)$, and $U(1)_X$, 
respectively\footnote{We note in passing that it is simple to extend the gauge symmetry to gauge also lepton number: $SU(4) \times SU(2)_L \times U(1)_X \times U(1)_L$. 
The new anomalies are cancelled by three families of right-handed neutrinos:
\bea
&&\hspace{-4mm}\hat{Q}_{L} = \left(4, 2, 0, 0\right) , \ \ \ \ \ \ \, \ \hat{u}_{R} = \left(4, 1, \tfrac{1}{2}, 0\right) , \ \  \ \ \hat{d}_{R} = \left(4, 1, -\tfrac{1}{2}, 0\right)  ,  \nonumber\\
&&\hspace{-4mm}Q'_{R} = \left(1, 2, -\tfrac{1}{2}, 1\right)  , \ \ \ u'_{L} = \left(1, 1, 0, 1\right)  , \ \ \ \ \ \, d'_{L} = \left(1, 1, -1, 1\right) ,\nonumber\\
&&\hspace{-4mm}\ l_{L} \,\,= \left(1, 2, -\tfrac{1}{2}, 1\right)  , \ \ \ e_{R} = \left(1, 1, -1, 1\right) ,\ \ \ \nu_R = (1, 1, 0, 1) \ , \nonumber\\
&&\hspace{-4mm} \,S_L = (1, 1, 0, -2)\, , \ \ \ \  \ \ \hat{\Phi}  = \left(4, 1, \tfrac{1}{2}, -1\right)  , \ \ \  H= \left(1, 2, \tfrac{1}{2}, 0\right),
\eea
where $S_L$ is the additional Higgs needed to break $U(1)_L$.
This content allows for new Yukawa terms producing Dirac masses for the neutrinos and a Majorana 
mass term arising from $S_L \nu_R \nu_R$, thus accommodating a type I seesaw mechanism for neutrino masses.}.

The scalar sector contains the $SU(4)$ quadruplet $\hat{\Phi}$, whose vacuum expectation value (VEV),
\bea
\langle\hat\Phi\rangle = \tfrac{1}{\sqrt2}\left(0 \ \ 0 \ \  0 \ \ V\right)^T,
\eea
breaks the gauge symmetry down to the SM.  Hypercharge
emerges as a combination of the $T^{15}$ generator of $SU(4)$ and $U(1)_X$,
\bea
Y =   X +  \sqrt{\frac{2}{3}} ~ T^{15} ~.
                    \label{111-3}
\eea
The $SU(2)_L$ doublet Higgs $H$ breaks the electroweak symmetry down to electromagnetism as usual.

The $SU(4)$ breaking results in seven massive gauge bosons which organize themselves into
three complex vector fields ${G'}^\alpha_{\!\!\mu}$ transforming as a color triplet with mass
\bea
m_{G'} = \tfrac{1}{2}\,g_4 \, V  \, ,
\eea
mediating interactions between each SM quark and its partner:
\bea
\frac{~g_4}{\sqrt{2}} \left\{
\bar{{Q}}_{L}^{\alpha}\,{\slashed{G}\,\!'}^\alpha \tilde{Q}_L 
+  \bar{{u}}_{R}^{\alpha}\,{\slashed{G}\,\!'}^\alpha \tilde{u}_R 
+  \bar{{d}}_{R}^{\alpha}\,{\slashed{G}\,\!'}^\alpha \tilde{d}_R \right\}  + {\rm h.c.}\ ;
\eea
and a neutral $Z'$ gauge boson with mass
 \bea
m_{Z^\prime} = \tfrac{1}{2}\sqrt{g_X^2+ \tfrac{3}{2} g_4^2} \ \ V ~,
\eea
which couples to pairs of quarks, quark partners or leptons with strength:
\bea
-\frac{g_Y}{\tan \theta_4}
\left[  \sqrt{\frac{2}{3}} T^{15} -  X  \tan^2 \!\theta_4 \right].
\label{Zcouplings}
\eea
The angle in $\sin{\theta_4} \equiv g_X / \sqrt{g_X^2+ 3g_4^2/2}$ can be determined based on the hypercharge and strong couplings at scale $V$.  For $V \sim $~TeV
it is predicted that $\sin \theta_4 \approx 0.28$.

\subsection{Quark Partner Masses and MFV}

The masses of the quark partners receive contributions from the VEVs of both $\hat{\Phi}$ and $H$ via Yukawa interactions:
\bea
 & &\hspace{-5mm} Y_Q^{ab}\,\bar{\hat{Q}}_L^a \,\hat{\Phi}\, {Q'}_{\!\!R}^{b}
 + Y_u^{ab}\,\bar{\hat{u}}_R^a \, \hat{\Phi}\, {u'}_{\!\!L}^{b}
 + \ Y_d^{ab} \, \bar{\hat{d}}_R^a \,\hat{\Phi}\,{d'}_{\!\!L}^{b} \nonumber \\   
&&\hspace{-5mm} +  \  y_u^{ab}\, \bar{\hat{Q}}_L^a\, \tilde{H} \,\hat{u}_R^b + y_d^{ab}\, \bar{\hat{Q}}_L^a \,H \,\hat{d}_R^{\,b} \nonumber\\ 
&&\hspace{-5mm} + \  {{y'_u}}^{\!ab}\, \bar{Q'}_{\!\!R}^a\,\tilde{H} \, {u'}_{\!\!L}^b + {{y'_d}}^{\!ab}\, \bar{Q'}_{\!\!R}^a \,H \,{d'}_{\!\!L}^b
+ {\rm h.c.} \ ,
\eea
where the $Y$ couplings marry the quark partners to the spectator fields $Q'$, $u'$, and $d'$;
the $y$ couplings contain the SM Yukawa interactions for the quarks, and the $y^\prime$ couplings
lead to mixing between the quark partner singlets and doublets.
The result, denoting $\tilde{Q} = (\tilde{U}, \tilde{D})$ and so on for $Q'$, is a pair of $6 \times 6$
matrices,
\bea
& & \hspace*{-0.75cm}
\frac{1}{\sqrt{2}}
\left(\!
\overline{\tilde{U}}_{\!L} ~~ \overline{u}^\prime_L
\right)
 \left(\!
  \begin{array}{cc}
    Y_Q V & y_u v \\
    \left( y^\prime_u v \right)^\dagger & \left( Y_u V \right)^\dagger \\
  \end{array}\!
\right)
\left(\!\!
  \begin{array}{c}
   U^\prime_R \\
    \tilde{u}_R \\
  \end{array}\!\!
\right) \nonumber \\
& & \hspace*{-0.75cm} + \frac{1}{\sqrt{2}}
\left(\!
\overline{\tilde{D}}_L ~~ \overline{d}^\prime_L
\right)
 \left(\!
  \begin{array}{cc}
    Y_Q V & y_d \,v \\
    \left( y^\prime_d v \right)^\dagger & \left( Y_d V \right)^\dagger \\
  \end{array}\!
\right)
\left(\!\!
  \begin{array}{c}
   D^\prime_R \\
    \tilde{d}_R \\
  \end{array}\!\!
\right)
+ {\rm h.c.} \ ,
\eea
where $v \simeq 246$~GeV is the SM Higgs VEV.  
The eigenvalues of those two matrices yield the 
masses of six electrically neutral states (combinations of $\tilde{u}$ and $\tilde{U}$)
and six electric charge minus one states (combinations of $\tilde{d}$ and $\tilde{D}$).

Under the SM flavor symmetries, $\tilde{Q}$, $\tilde{u}$, and $\tilde{d}$ each transform as triplets of
$SU(3)_Q$, $SU(3)_u$, and $SU(3)_d$, respectively.  The simplest 
choice\footnote{An alternate choice leads to $Y_Q \propto y_u$, $Y_d \propto y_d^\dagger$, and
$Y_u, y_u', y_d' \propto y_u^\dagger$, which would result in large hierarchies in the quark partner masses
and require couplings which are nonperturbative.} 
is to assign the spectator fields $Q'$, $u'$, and $d'$ to also transform as triplets under
$SU(3)_Q$, $SU(3)_u$, and $SU(3)_d$, respectively.  
MFV then dictates that, to leading order in the spurions
$y_u$ and $y_d$, the remaining Yukawa interactions are given by,
\bea
Y_Q^{ab} = Y_Q ~\mathbb{1}, ~~~~ Y_u^{ab} = Y_u ~\mathbb{1}, ~~~ Y_d^{ab} = Y_d ~\mathbb{1} \ ,
\eea
where $\mathbb{1}$ denotes the $3 \times 3$ unit matrix, and,
\bea
y_u^\prime = \eta ~ y_u\ , ~~~~~~~ y_d^\prime = \eta^\prime ~y_d\ .
\eea
After imposing MFV, the masses of the quark partners are determined by the five parameters:
$Y_Q$,  $Y_u$,  $Y_d$, 
$\eta$, and $\eta^\prime$, in terms of the SM flavor structure encoded
in $y_u$ and $y_d$.

In the SM quark mass basis, the mass matrices for the partners take the block form,
\bea
& & \hspace*{-0.75cm}
\left(\!
\overline{\tilde{U}}_{\!L} ~~ \overline{u}^\prime_L
\right)
 \left(\!
  \begin{array}{cc}
    M~ \mathbb{1} & m_u \\
    \eta ~m_u & m ~\mathbb{1} \\
  \end{array}\!
\right)
\left(\!\!
  \begin{array}{c}
   U^\prime_R \\
    \tilde{u}_R \\
  \end{array}\!\!
\right)  \nonumber \\
& & \hspace*{-0.75cm}+ 
\left(\!
\overline{\tilde{D}}_L ~~ \overline{d}^\prime_L
\right)
 \left(\!
  \begin{array}{cc}
    M ~\mathbb{1} & m_d \\
    \eta'\,m_d & m^\prime ~\mathbb{1} \\
  \end{array}\!
\right)
\left(\!\!
  \begin{array}{c}
   D^\prime_R \\
    \tilde{d}_R \\
  \end{array}\!\!
\right)
+ {\rm h.c.} \ ,
\eea
where $m_u$ and $m_d$ are diagonal $3 \times 3$ matrices whose entries are the up-type and down-type SM quark masses,
whereas $M \equiv Y_Q V / \sqrt{2}$, \ $m \equiv Y_u V / \sqrt{2}$ \  and \ $m^\prime \equiv Y_d V / \sqrt{2}$.  

To good 
approximation (assuming $\eta'$ is not extremely large)
the partners of the first and second generation quarks
consist of
two degenerate $SU(2)$ doublets of mass $\simeq M$ (along with the partner of the left-handed bottom quark),
three degenerate charge $-1$ singlet states of mass $m^\prime$, and two degenerate neutral singlet
states of mass $m$, with tiny
intergenerational mixing and thus negligible contributions to 
$K$--$\overline{K}$, $B$--$\overline{B}$, and $D$--$\overline{D}$ mixing.

The large top mass results in non-negligible mixing between the $SU(2)$ singlet and doublet top partners, so that
their masses are split from $M$ and $m$.
The lighter of the two states, which we denote as $\chi$,
is stable due to a global $U(1)$ symmetry left over after the $SU(4)$ breaking and plays the role of DM.
Its couplings to the $W$ and $Z$ bosons are controlled by the admixture of the $SU(2)$ doublet, which in turn is controlled by $M$, $m$, and $\eta$.
The mass and gauge eigenstates are related by two mixing angles,
\bea
\chi_L &=&  \cos \theta_L \ t'_L + \sin \theta_L \ \tilde{T}_L  \ , \\
\chi_R &=& \cos \theta_R \ \tilde{t}_R +\sin \theta_R \ T'_R \ .
\eea
In the limit $M \gg m, m_t$\,,
\bea
m_{\chi}  & \simeq & m - \frac{m_t^2}{M} \eta ~,
\eea
\bea
\sin \theta_R   \simeq  - \frac{m_t}{M} \eta ~, ~~~~~ \sin \theta_L  \simeq  - \frac{m_t}{M} ~.
\eea

As shown below, to evade strong constraints from searches for DM scattering with nuclei, the singlet component should be 
dominant, i.e. $\theta_R, \theta_L \ll 1$.
To simplify our parameter space, we consider $\eta = 1$, for which the two mixing angles are the same.  We
parameterize the degree of the $SU(2)$ doublet inside $\chi$ by 
\bea
\epsilon \equiv \sin \theta_R = \sin \theta_L \ll 1~.
\eea

\section{Scattering and Annihilation}
\label{sec:dm}

In this section we estimate the cross sections for $\chi$ to scatter with heavy nuclei or annihilate into SM states.

\subsection{Direct Detection}

     \tikzset{
      my box/.style = {draw, minimum width = 2em, minimum height=1em},
particle/.style={thick,draw=black, postaction={decorate},
    decoration={markings,mark=at position .65 with {\arrow[black]{triangle 45}}}},
particle2/.style={thick,draw=black, postaction={decorate},
    decoration={markings,mark=at position .55 with {\arrow[black]{triangle 45 reversed}}}},
    particle22/.style={thick,draw=black, postaction={decorate},
    decoration={markings,mark=at position .6 with {\arrow[black]{triangle 45 reversed}}}},
        particle222/.style={thick,draw=black, postaction={decorate},
    decoration={markings,mark=at position .6 with {\arrow[black]{triangle 45}}}},
            particle223/.style={thick,draw=black, postaction={decorate},
    decoration={markings,mark=at position .7 with {\arrow[black]{triangle 45}}}},
particle3/.style={thick,draw=black, postaction={decorate},
    decoration={markings,mark=at position .35 with {\arrow[black]{triangle 45}},mark=at position .88 with {\arrow[black]{triangle 45}}}},
gluon/.style={ultra thick, decorate, draw=black,
    decoration={coil,aspect=0}},
 gluon3/.style={thick, decorate, draw=black,
    decoration={coil,aspect=0}},  
gluon4/.style={decorate, draw=black, thick,
    decoration={coil,amplitude=4pt, segment length=5pt}},
Higgs/.style={thick, dashed, draw=black}
    }
    
\begin{figure}[t!]
\begin{tikzpicture}[node distance=0.6cm and 0.9cm]
\coordinate[label=left:$\mathlarger{\mathlarger{{\chi}}}$] (e1);
\coordinate[below right=of e1] (aux1);
\coordinate[above right=of aux1,label=right:$\mathlarger{\mathlarger{{\chi}}}$] (e2);
\coordinate[below=0.9cm of aux1] (aux2);
\coordinate[below left=of aux2,label=left:$\mathlarger{\mathlarger{{q}}}$] (e3);
\coordinate[below right=of aux2,label=right:$\mathlarger{\mathlarger{{{q}}}}$] (e4);
\draw[particle] (e1) -- (aux1);
\draw[particle223] (aux1) -- (e2);
\draw[particle] (e3) -- (aux2);
\draw[particle223] (aux2) -- (e4);
\draw[gluon] (aux1) -- node[label=right:$\mathlarger{\mathlarger{ \ Z}'}$] {} (aux2);
\node[below] at (current bounding box.south){};
\end{tikzpicture}\hspace*{\fill}
\begin{tikzpicture}[node distance=0.6cm and 0.9cm]
\coordinate[label=left:$\mathlarger{\mathlarger{{\chi}}}$] (e1);
\coordinate[below right=of e1] (aux1);
\coordinate[above right=of aux1,label=right:$\mathlarger{\mathlarger{{\chi}}}$] (e2);
\coordinate[below=0.9cm of aux1] (aux2);
\coordinate[below left=of aux2,label=left:$\mathlarger{\mathlarger{{q}}}$] (e3);
\coordinate[below right=of aux2,label=right:$\mathlarger{\mathlarger{{{q}}}}$] (e4);
\draw[particle] (e1) -- (aux1);
\draw[particle223] (aux1) -- (e2);
\draw[particle] (e3) -- (aux2);
\draw[particle223] (aux2) -- (e4);
\draw[gluon3] (aux1) -- node[label=right:$\mathlarger{\mathlarger{ \ Z}}$] {} (aux2);
\node[below] at (current bounding box.south){};% {label fig one};
\end{tikzpicture}\hspace*{\fill}
\begin{tikzpicture}[node distance=0.6cm and 0.9cm]
\coordinate[label=left:$\mathlarger{\mathlarger{{\chi}}}$] (e1);
\coordinate[below right=of e1] (aux1);
\coordinate[above right=of aux1,label=right:$\mathlarger{\mathlarger{{\chi}}}$] (e2);
\coordinate[below=0.9cm of aux1] (aux2);
\coordinate[below left=of aux2,label=left:$\mathlarger{\mathlarger{{q}}}$] (e3);
\coordinate[below right=of aux2,label=right:$\mathlarger{\mathlarger{{{q}}}}$] (e4);
\draw[particle] (e1) -- (aux1);
\draw[particle223] (aux1) -- (e2);
\draw[particle] (e3) -- (aux2);
\draw[particle223] (aux2) -- (e4);
\draw[Higgs] (aux1) -- node[label=right:$\mathlarger{\mathlarger{ \ h}}$] {} (aux2);
\node[below] at (current bounding box.south){};% {label fig one};
\end{tikzpicture}\hspace*{\fill}
\caption{\small{Representative Feynman diagrams contributing  to $\chi$ \hspace{20mm} {\textcolor{white}{..}} \hspace{-28mm} interacting with quarks.}}
\label{fig:1}
\end{figure}
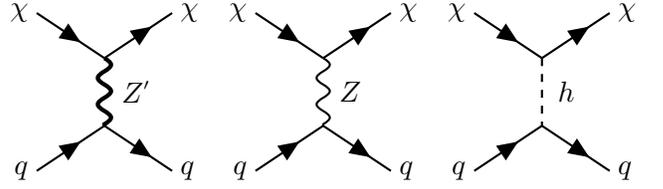       
 
The DM particle $\chi$ interacts with quarks either through the exchange of a heavy $Z'$  or (via its doublet component admixture)
the electroweak $Z$ and the SM Higgs boson (see Fig.~\ref{fig:1}).
In the limit of nonrelativistic $\chi$, the effective Lagrangian relevant for the spin-independent cross section can be written as: 
\bea
\mathcal{L}_{\rm eff} = \sum_{q} \Big[\,c_1^{(q)}\, (\bar{\chi}  \,\chi) \left(\bar{q}\,q\right)  
+  c_2^{(q)} \,(\bar{\chi}  \gamma^\mu \chi) \left(\bar{q} \gamma_\mu q\right)\Big],
\label{list}
\eea
where the coefficients $c_{1,2}^{(q)}$ are given by:
\bea
c_1^{(q)} &=& -\frac{\epsilon}{2} \frac{y_t \,y_q}{m_h^2} \hspace{55mm} \  \nonumber\\
c_2^{(u,c)} &=& -\frac{g_Y^2}{48}  \Bigg[ \frac{1}{m_{Z'}^2} \left(\frac{1+\tan^2 \theta_4}{\tan^2\theta_4}\right)\left(2-3\tan^2\theta_4\right)\nonumber\\
&&\hspace{6mm}-\frac{\epsilon^2}{m_Z^2}\left(\frac{2}{\sin^2\theta_W}\right)\left(3-8\sin^2\theta_W\right)\Bigg] ,\nonumber\\
c_2^{(d,s,b)} \hspace{-1mm}&=& -\frac{g_Y^2}{48} \Bigg[ \frac{1}{m_{Z'}^2} \left(\frac{1+\tan^2 \theta_4}{\tan^2\theta_4}\right)\left(2+3\tan^2\theta_4\right)\nonumber\\
&&\hspace{4mm}-\frac{\epsilon^2}{m_Z^2}\left(\frac{2}{\sin^2\theta_W}\right)\left(-3+4\sin^2\theta_W\right)\Bigg],\nonumber\\
c_2^{(t)} &\hspace{-3mm}=&\hspace{-2mm} -\frac{g_Y^2}{48}  \Bigg[ \frac{1}{m_{Z'}^2} \left(\frac{1+\tan^2 \theta_4}{\tan^2\theta_4}\right)\left(2-3\tan^2\theta_4\right)\nonumber\\
&&\hspace{-10mm}-\frac{\epsilon^2}{m_Z^2}\left(\frac{2}{\sin^2\theta_W}\right)\left(3-8\sin^2\theta_W\right)\Bigg]-\frac{g_3^2}{8}\frac{1}{m_{G'}^2} .
\eea 

Equation (\ref{list}) maps onto effective interactions between $\chi$ and the nucleon $N=\{ p, n \}$:
\bea
\hspace{-2mm}\mathcal{L}_{\rm eff} = \hspace{-2mm}\sum_{N=p,n} \hspace{-1mm}C_1^{(N)} (\bar{\chi}  \,\chi) \left(\bar{N}\,\!N\right)  
+  C_2^{(N)} (\bar{\chi}  \gamma^\mu \chi) \left(\bar{N} \gamma_\mu N\right) 
\eea
where (e.g. \cite{delnobile}),
\bea
\ C_1^{(N)} \!&=&\! \sum_{q=u,d,s} c_1^{(q)} \frac{m_N}{m_q} f_q^{(N)} + \frac{2}{27}\, f^{(N)}_G \!\!\!\sum_{q=c,b,t}c_1^{(q)}\frac{m_N}{m_q}\ , \nonumber\\
C_2^{(p)} &=& 2\,c_2^{(u)}+ c_2^{(d)}, \ \ \ \ C_2^{(n)} = c_2^{(u)}+ 2\,c_2^{(d)}, 
\eea
with the coefficients  \cite{gondolo}:
\bea
&&\hspace{-0mm}f_u^{(p)} \!=\! 0.023, \, \ f_u^{(n)} \!=\! 0.018, \, \  f_d^{(p)} \!=\! 0.033, \, \  f_d^{(n)} \!=\! 0.042, \nonumber\\
&& f_s^{(p)} = f_s^{(n)}  = 0.26, \, \ f_G^{(p)} = 0.684, \, \ f_G^{(n)} = 0.68\ .
\eea
The zero-velocity spin-independent cross section for $\chi$ to scatter with a nucleon is thus:
\bea
\hspace{-10mm}\sigma_{\rm SI}   &=&  \frac{1}{\pi} \frac{m_{\chi}^2 \,m_N^2}{(m_{\chi} + m_N)^2} \frac{1}{A^2} \nonumber\\
& \hspace{-7mm}\times& \hspace{-5mm}\bigg[\,Z\left(C_1^{(p)} + C_2^{(p)}\right) +(A-Z)\left(C_1^{(n)} + C_2^{(n)}\right)\bigg]^2 \!.
\label{si}
\eea

Currently, the most stringent limit on $\sigma_{\rm SI}$ for heavy DM comes from the LUX experiment \cite{limit}.
For $m_\chi$ much larger than the mass of a xenon atom, the limit scales simply as $\propto m_\chi$, reflecting
the fact that for constant local DM energy density the number density falls as $\propto 1 / m_\chi$.
Neglecting the subdominant Higgs contribution, for large masses the LUX limit imposes a constraint on $V$ and $\epsilon$:
\bea
 \left[ \left( \frac{7 {\rm \ TeV}}{V}\right)^2 
  %+ 8 \, \epsilon
 + 450 \,\epsilon^2\, \right]^2
  \lesssim \ \frac{m_{\chi}}{1 {\rm \ TeV}} \,,
\label{b}
\eea
generally requiring $V \gtrsim 7$~TeV and $\epsilon  \lesssim 0.05$.  The bounds on $V$ from LUX
are typically stronger than those imposed by null searches for $Z^\prime$ bosons at the LHC
\cite{Fornal:2015boa,Carena} or precision electroweak constraints.

\subsection{Annihilation}

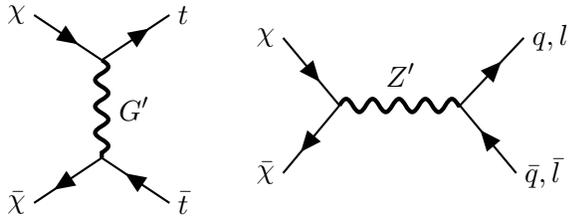
\begin{figure}[t!]
\hspace*{\fill}
\begin{tikzpicture}[node distance=0.6cm and 0.9cm]
\coordinate[label=left:$\mathlarger{\mathlarger{{\chi}}}$] (e1);
\coordinate[below right=of e1] (aux1);
\coordinate[above right=of aux1,label=right:$\mathlarger{\mathlarger{{t}}}$] (e2);
\coordinate[below=1.3cm of aux1] (aux2);
\coordinate[below left=of aux2,label=left:$\mathlarger{\mathlarger{\bar{\chi}}}$] (e3);
\coordinate[below right=of aux2,label=right:$\mathlarger{\mathlarger{\bar{{t}}}}$] (e4);
\draw[particle] (e1) -- (aux1);
\draw[particle223] (aux1) -- (e2);
\draw[particle2] (e3) -- (aux2);
\draw[particle22] (aux2) -- (e4);
\draw[gluon] (aux1) -- node[label=right:$\mathlarger{\mathlarger{ \ G'}}$] {} (aux2);
\end{tikzpicture}\hspace*{\fill}
\begin{tikzpicture}[node distance=0.9cm and 0.75cm]
\coordinate[label=left:$\mathlarger{\mathlarger{\chi}}$] (e1);
\coordinate[below right=of e1] (aux1);
\coordinate[above=1cm,right=3.2cm,label=right:$\mathlarger{\mathlarger{{q, l \ \ \ }}}$] (e2);
\coordinate[right=1.6cm of aux1] (aux2);
\coordinate[below left=of aux1,label=left:$\mathlarger{\mathlarger{\bar{\chi}}}$] (e3);
\coordinate[below right=of aux2,label=right:$\mathlarger{\mathlarger{\bar{{q}},\bar{l}}}$] (e4);
\draw[particle] (e1) -- (aux1);
\draw[particle] (aux2) -- (e2);
\draw[particle2] (e3) -- (aux1);
\draw[particle22] (aux2) -- (e4);
\draw[gluon] (aux1) -- (aux2);
\node [label={[label distance=1.16cm]-10:$\mathlarger{\mathlarger{{{Z'}}}}$}] {};
\node[below] at (current bounding box.south){ \ \ \color{white} \footnotesize{ n } };
\end{tikzpicture}
\caption{\small{Representative Feynman diagrams for $\chi \bar{\chi}$ annihilation \hspace{20mm} {\textcolor{white}{..}} \hspace{-24.5mm}into SM quarks and leptons.}}
\label{fig:2}
\end{figure}

Due to the severe constraint on the electroweak doublet admixture $\epsilon$ in Eq.~(\ref{b}), the DM
annihilation into electroweak and Higgs bosons is highly suppressed.  The dominant annihilation channels
are SM quarks and leptons, mediated by the heavy gauge bosons $G^\prime$ and $Z^\prime$, as
shown in Fig.~\ref{fig:2}.  Since the couplings are essentially fixed by the embedding of $SU(3)_c$ and $U(1)_Y$, and the
masses are related to one another by the $SU(4)$ breaking scale $V$, if one assumes a standard thermal history of the Universe
the resulting relic density of $\chi$ through freeze-out is determined by the values of $V$ and $m_\chi$ 
with little other model dependence.

\begin{figure}[t!]
\centering
\includegraphics[scale=0.212, trim={0cm 0mm 0cm -2mm}]{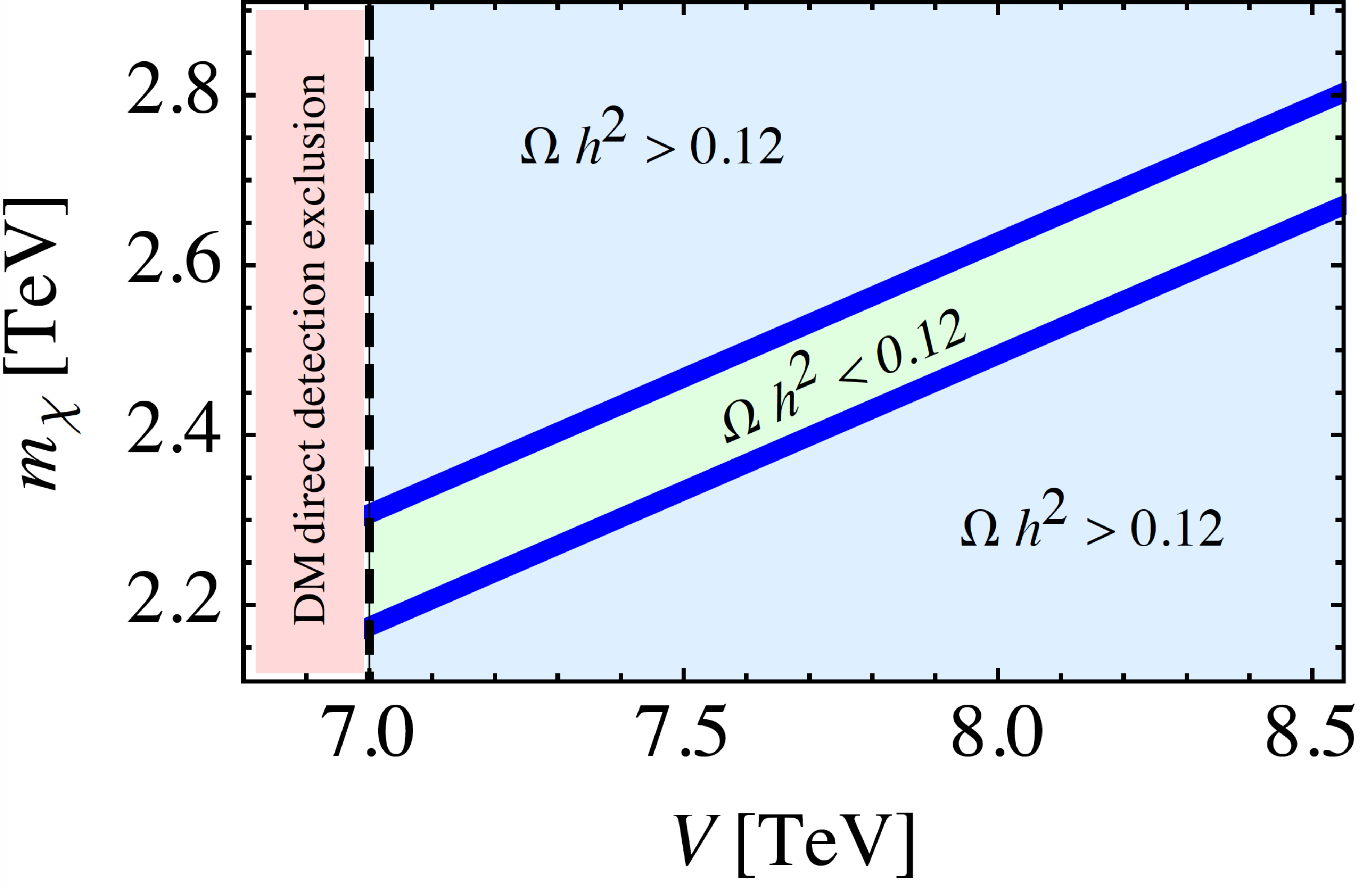}
\caption{\small{The relic density of $\chi$ relative to the Planck preferred 
value {\textcolor{white}{.............}}  \cite{Ade}    for a standard cosmological history, with
the blue lines saturating $\Omega h^2 \simeq 0.12$, in the plane of $V$ and $m_\chi$.}}  
\label{fig:3}
\end{figure}

The resulting relic density is shown in Fig.~\ref{fig:3}.  Cross sections of order $\sim 3 \times 10^{26}$~cm$^3$/s 
are obtained only
when $m_\chi$ and $V$ are chosen such that annihilation is modestly enhanced by the $Z^\prime$ pole, which happens for
$m_{\chi} \approx V/3$.  As is usual in such cases, for fixed $V$ (and thus $m_{Z'}$) there are two values of $m_\chi$ for which
the thermal relic density is saturated on either side of $m_{Z'}$.  Between those two values
of $m_\chi$ the cross section is larger and the relic density is typically too small, whereas outside of this range the cross section is too
small, and the relic density is in general too large.  Limits from indirect detection for this mass range are typically too weak by
a few orders of magnitude to provide useful constraints on this parameter space \cite{Acciari:2010ab,Abramowski:2014tra}.

\vspace{5mm}

\section{Collider Phenomenology}
\label{sec:lhc}

Given the flavorful nature of the DM, signatures at high energy colliders are typically rich in top quarks.  Because of  its 
relatively large coupling to SM quarks and leptons, prospects to observe the $Z^\prime$ at run II of the LHC are
good \cite{Fornal:2015boa}, though connecting it to a theory of DM will be more of a challenge.  Even for a standard
thermal relic, the DM could be heavy enough that there will be no on-shell $Z^\prime$ decays into it, and even if there are
$Z^\prime \rightarrow \chi \,\bar{\chi}$ decays open, identifying them is likely to prove challenging.  It may fall to future high energy
colliders to establish the connection between the $Z^\prime$ and DM.

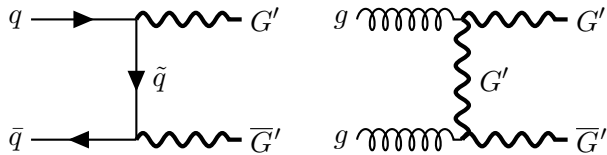
\begin{figure}[t!]
\vspace{5mm}
\hspace{1mm}
\begin{tikzpicture}[node distance=0.0cm and 1.4cm]
\coordinate[label=left:$\mathlarger{\mathlarger{q}}$] (e1);
\coordinate[below right=of e1] (aux1);
\coordinate[above right=of aux1] (e2);
\coordinate[below=1.6cm of aux1] (aux2);
\coordinate[below left=of aux2,label=left:$\mathlarger{\mathlarger{\bar{q}}}$] (e3);
\coordinate[below right=of aux2] (e4);
\draw[particle222] (e1) -- (aux1);
\draw[gluon] (aux1) -- (e2);
\coordinate[above=1.5cm,right=2.8cm,label=right:$\mathlarger{\mathlarger{G'}}$] (e2);
\draw[particle2] (e3) -- (aux2);
\draw[gluon] (aux2) -- (e4);
\coordinate[below right=of aux2,label=right:$\mathlarger{\mathlarger{\overline{G}\hspace{0.2mm}'}}$] (e4);
\draw[particle222] (aux1) -- node[label=right:$\mathlarger{\mathlarger{ \ \tilde{q}}}$] {} (aux2);
\end{tikzpicture}\hspace{4mm}
\begin{tikzpicture}[node distance=0.0cm and 1.4cm]
\coordinate[label=left:$\mathlarger{\mathlarger{g}}$] (e1);
\coordinate[below right=of e1] (aux1);
\coordinate[above right=of aux1] (e2);
\coordinate[below=1.6cm of aux1] (aux2);
\coordinate[below left=of aux2,label=left:$\mathlarger{\mathlarger{g}}$] (e3);
\coordinate[below right=of aux2] (e4);
\draw[gluon4] (e1) --  (aux1);
\coordinate[above=1.5cm,right=2.8cm,label=right:$\mathlarger{\mathlarger{G'}}$] (e2);
\draw[gluon4] (e3) -- (aux2);
\draw[gluon] (aux2) -- (e4);
\draw[gluon] (aux1) -- (e2);
\coordinate[below right=of aux2,label=right:$\mathlarger{\mathlarger{\overline{G}\hspace{0.2mm}'}}$] (e4);
\draw[gluon] (aux1) -- node[label=right:$\mathlarger{\mathlarger{ \ G'}}$] {} (aux2);
\end{tikzpicture}
\caption{\small{Representative Feynman diagrams for pair production \hspace{20mm} {\textcolor{white}{..}} \hspace{-24mm} of $G' \overline{G}\hspace{0.2mm}'$ at
a hadron collider.}}
\label{fig:4}
\end{figure}

At a future 100 TeV hadron collider, one of the 
most relevant signatures is pair production of the colored $G'$ from either a $gg$ or $q \bar{q}$ partonic initial state
(Fig.~\ref{fig:4}).  The $G'$ will decay into a quark plus a quark partner,
resulting in signatures with jets and missing transverse momentum.  In particular, decay into $\chi$ plus a top quark results in
$t + {\rm MET}$, or cascading through one of the other quark partners results in $t + j + {\rm MET}$.  For pair production, the
signatures are:
\begin{itemize}
\item[{{(a)}}] $\ \ p \, p \,\rightarrow\, G' \,\overline{G}\hspace{0.2mm}'\rightarrow\, t \, \bar{t} \,+\, {\rm MET}$\ ,
\item[{(b)}] $\ \ p \, p \,\rightarrow\, G' \,\overline{G}\hspace{0.2mm}' \rightarrow\, 2\,j \,+\,t \, \bar{t} \,+\, {\rm MET}$~,
\end{itemize}
where $j$ denotes a light (unflavored or $b$ jet).
These signatures are similar to the scalar top or gluino ones from supersymmetry, with mild differences caused by the different spins of
the produced particles.

We estimate the rate for pair production of $G'$ by implementing its couplings into a FeynRules model  \cite{Feyn,Degrande:2011ua}, 
which is used by MadGraph \cite{Alwall:2014hca} to compute the inclusive cross section.  The $gg$ initiated process has a rate
determined entirely by gauge invariance under $SU(3)_c$, and thus the only model dependence is the mass of the $G'$ itself.  The quark
initiated process proceeds via exchange of the quark partners, and thus is sensitive to their mass spectrum.  We  fix this
spectrum by choosing $m = m'$ such that the DM is a canonical thermal relic (see Fig.~\ref{fig:3}) and $\epsilon \ll 1$
(which requires $M$ to be large enough such that the left-handed quark partners are largely irrelevant).

\begin{figure}[t!]
\centering
\includegraphics[scale=0.225, trim={0cm 0mm 0cm 0mm}]{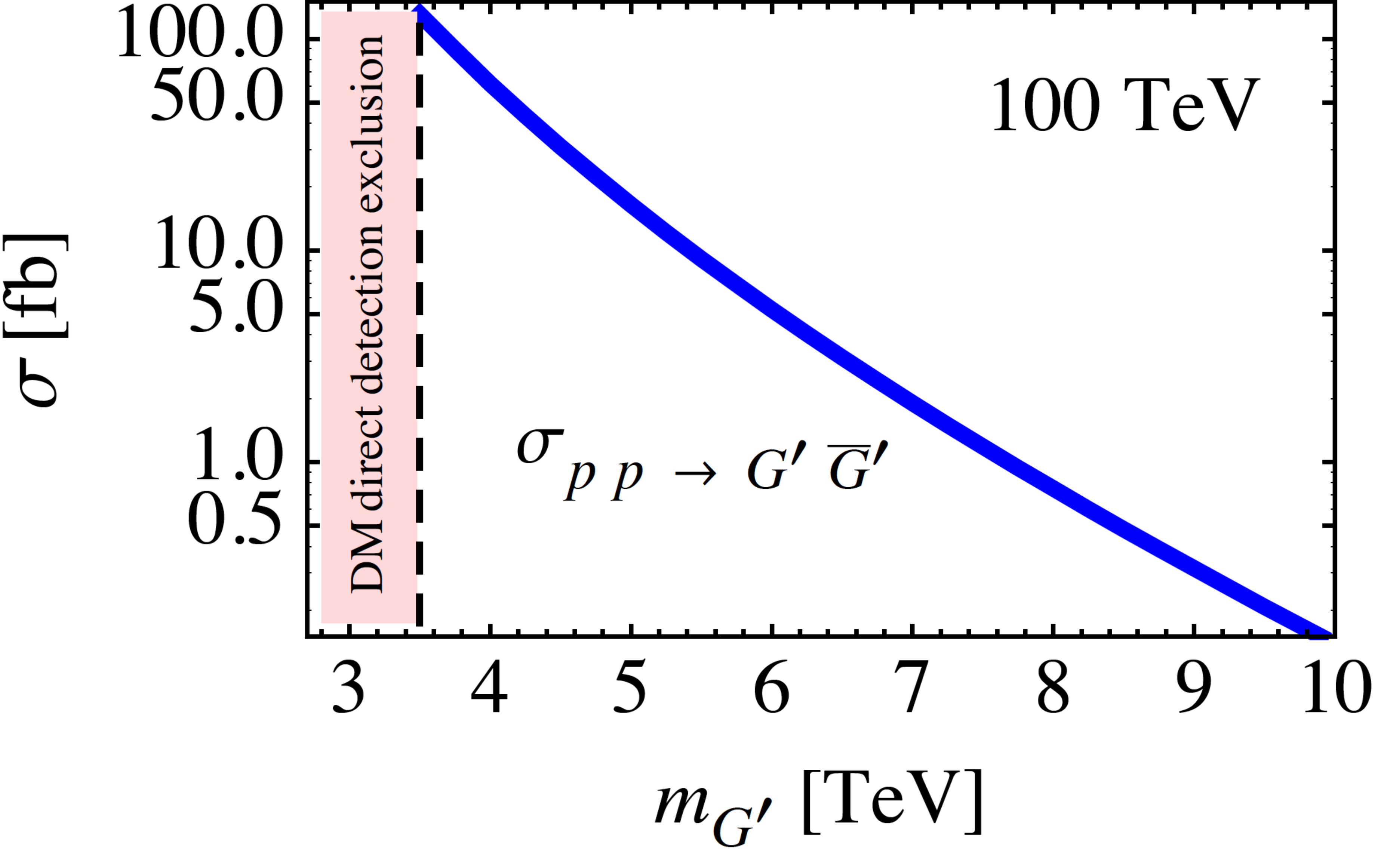}
\caption{\small{Cross section for $\ p \, p \,\rightarrow\,G' \,\overline{G}\hspace{0.2mm}'\,$ as a function of the \hspace{20mm} {\textcolor{white}{..}} \hspace{-2.5mm} $G'$ gauge boson mass for $E_{\rm CM} = 100 \ {\rm TeV}$.}}  
\label{fig:5}
\end{figure}

The resulting cross section is shown in Fig.~\ref{fig:5} as a function of the $G'$ mass.  For the quark partner masses used to
generate this plot, the branching ratio for $G' \rightarrow \chi + t$ will be about $1/6$, whereas that for
$G' \rightarrow \chi + j + t$ is around $5/6$.  The backgrounds for signals such as these at a 100 TeV collider are estimated to be on
the order of a femtobarn \cite{Low:2014cba,Cohen:2014hxa}
(after cuts for which the signal events should pass with reasonable efficiency), 
indicating that $G'$ masses on the order of $\sim 7.5$~TeV can be probed by this facility.
Observation of the $G'$ bosons would be the real clue as to the underlying $SU(4)$ gauge symmetry and its connection
to DM.

\section{Conclusions}
\label{sec:conclusions}

We have analyzed a novel extension of the Standard Model in which color is unified with baryon number into a single $SU(4)$ gauge group. 
The theory contains the minimal number of new degrees of freedom consistent with the enlarged gauge symmetry and includes
a dark matter candidate.  Constraints from existing searches at colliders and direct detection experiments, along with the dark matter relic abundance,
suggest $\sim$~TeV scale masses for these new particles, setting a lower bound on the $SU(4)$ breaking scale of 
$\sim 7$ TeV and point to the dark matter being mostly an electroweak singlet. 

LHC searches are currently not very constraining, but ultimately have good prospects to detect the $Z'$, which has large 
coupling to both quarks and leptons.  However, its connection to dark matter and the underlying $SU(4)$ symmetry are challenging
at the LHC, and would benefit greatly from searches at future colliders, including the 100 TeV $pp$ machine currently under discussion.

\subsection*{Acknowledgments}
We thank Arvind Rajaraman for discussions at the initial stages of the project
and to Liantao Wang and Mathew Low for discussion of their work \cite{Low:2014cba}. 
B.F. also acknowledges the enlightening conversations with Mark Wise, 
Clifford Cheung, Ann Nelson and David B. Kaplan.   
This research was supported in part by NSF grant PHY-1316792.

\end{document}